\RequirePackage{fix-cm}
\documentclass[pdftex,twocolumn,epjc3,final]{svjour3}  
\smartqed  
\RequirePackage{graphicx}
\RequirePackage{amsmath}
\RequirePackage{esvect}
\RequirePackage{lineno}
\RequirePackage{multirow}

\RequirePackage[colorlinks,citecolor=blue,urlcolor=blue,linkcolor=blue]{hyperref}
\journalname{Eur. Phys. J. A}

\begin{document}
\title{Double deeply virtual Compton scattering with positron beams at SoLID}

\titlerunning{pDDVCS-BCA@SoLID}   

\author{
S.~Zhao\thanksref{addr1} 
\and
A.~Camsonne\thanksref{addr2} 
\and
D.~Marchand\thanksref{addr1} 
\and
M.~Mazouz\thanksref{addr5} 
\and
N.~Sparveris\thanksref{addr4} 
\and
S.~Stepanyan\thanksref{addr2} 
\and
E.~Voutier\thanksref{addr1} 
\and
Z.W.~Zhao\thanksref{addr3} 
}

\institute{Universit\'e Paris-Saclay, CNRS/IN2P3, IJCLab, 91405 Orsay, France \label{addr1}
\and
Duke University, Durham, NC 27708, USA \label{addr3}
\and
Thomas Jefferson National Accelerator Facility, Newport News, VA 23606, USA \label{addr2}
\and
Facult\'e des Sciences de Monastir, 5000 Monastir, Tunisia \label{addr5}
\and
Temple University, Philadelphia, PA 19122, USA \label{addr4}
}

\date{Draft : \today}

\maketitle

%
%

\hyphenation{Tho-mas}
\hyphenation{po-si-tron}
\hyphenation{mo-del}

%
%

\begin{abstract}

Double Deeply Virtual Compton Scattering (DDVCS) is the only experimental channel for the determination of the dependence of the Generalized Parton Distributions (GPDs) on both the average and the transferred momentum independently. The physics observables of the electron induced di-muon production reaction $\vv{e}^{\pm}p \to e^{\pm}p\mu^+\mu^-$ off unpolarized hydrogen are discussed. Their measurement with the high luminosity and large acceptance SoLID spectrometer at the Thomas Jefferson National Accelerator Facility, using polarized and unpolarized positron and electron beams at 11~GeV is investigated. This experimental configuration is shown to provide unprecedented access to the GPDs with the determination of the real and imaginary parts of the Compton Form Factor ${\mathcal H}$ in an unexplored phase space, and to enable an exploratory investigation of higher twist  effects.

\end{abstract}

%
%

\hyphenation{so-lid}
\hyphenation{ari-ses}
\hyphenation{Ca-lo-ri-me-ter}
\hyphenation{so-le-noid}

%
%

\section{Introduction}

The description of the partonic structure of hadronic matter through the Generalized Parton Distributions (GPDs) ~\cite{Mueller:1998fv} has profoundly extended the understanding of the structure and dynamics of the nucleon~\cite{Diehl:2003ny,BELITSKY20051}. Providing a link between electromagnetic form factors and parton distributions, the GPDs unify within the same formalism two different experimental expressions of the same physics reality {\it i.e.} the nucleon structure. The GPDs are the structure functions of the nucleon parameterizing the complex dynamics of partons inside the nucleon governed by the non-perturbative regime of Quantum Chromo-Dynamics. They provide a tomography of the nucleon from the correlations between transverse position and longitudinal momentum of partons~\cite{Burkardt:2000uu}. As a result of these position-momentum correlations, GPDs provide the ability to experimentally access the unknown orbital momentum contribution of partons to the total spin of the nucleon~\cite{Ji:1996ek}. They also enable indirect access to one of the gravitational form factors encoding the shear forces and pressure distribution on the partons in the proton~\cite{Polyakov:2002yz}. The GPDs appear as fundamental building elements of the nuclear structure knowledge, asking for a precise and complete experimental determination. 

The GPDs can be accessed in the hard scattering regime of exclusive processes, {\it i.e.} for high-enough virtuality $Q^2$ of the exchanged photon  and small-enough quadrimomentum transfer $t$ to the nucleon to allow the probe to couple to partons and ensure the factorization of the reaction amplitude. In addition to these variables, the GPDs also depend on the average longitudinal momentum fraction $x$ of the initial parton and on the transferred longitudinal momentum fraction $-2\xi$ to the final parton, the so-called the skewness. The privileged reaction for the GPDs mapping in this multi-dimensional space is the deeply virtual Compton scattering (DVCS) where the virtual photon generated by the lepton beam transformed into a real photon after  interacting with a parton from the nucleon~\cite{PhysRevLett.80.5064,Vanderhaeghen:1999xj}. At the leading order of the perturbation theory, the GPDs enter the cross section for this process through Compton form factors (CFFs) which imaginary part involved the GPDs at the $x$=$\pm \xi$ phase space points while the real part is the convolution integral of the GPDs and the parton propagators over the whole $x$ physics range~\cite{Belitsky:2001ns}. {\it In fine}, DVCS allows to investigate  unambiguously GPDs along the diagonals $x$=$\pm \xi$ and is therefore limited to a restricted region of the physics phase space. 

The strict Compton scattering of a virtual photon, in which the final photon remains virtual, has been suggested as a new reaction channel to overcome this limitation~\cite{PhysRevLett.90.012001,PhysRevLett.90.022001}. In this double deeply virtual Compton scattering (DDVCS) process, the virtuality of the final state photon indeed decouples the experimental $x$- and $\xi$-dependences opening off-diagonal investigation of the GPDs. However, the difficulty of the theoretical interpretation of the process $ep \to ep \bar{l}l$ when detecting the $e^+e^-$-pair from the decay of the final virtual photon, and the small magnitude of the cross section did forbid any reliable experimental study. 

The advent of the energy upgrade of the Continuous Electron Beam Accelerator Facility and the development of next generation large acceptance and high luminosity detection capabilities at the Thomas Jefferson National Accelerator Facility (JLab) procure ideal tools to overcome the previous limitations.  Specifically, the measurement of $\mu^+\mu^-$-pairs from the process ${\vv{e}}^{\pm} p \to e^{\pm} p \gamma^\ast \to e^{\pm}p \mu^+ \mu^-$, using the Solenoidal Large Intensity Device (SoLID)~\cite{Chen:2014psa} supplemented with a muon detector~\cite{Proposal1}, or using a modified CLAS12 spectrometer~\cite{Proposal2}, are unique opportunities for DDVCS investigations~\cite{Zhao:2020th}. In this process, the comparison between polarized electron and positron beams is, similarly to  DVCS~\cite{Voutier:2014kea}, essential to distinguish the different reaction amplitudes.

The present study investigates the perspectives of DDVCS measurements at SoLID with both electron and positron beams. The next section reviews the main characteristics of the DDVCS process and its benefits for the completion of the GPDs experimental program. The specificities of experimental observables for polarized lepton beams of opposite charge and their GPD content are further discussed in the following section, before addressing the description of the experimental configuration and the performance of possible measurements.

%
%

\section{Double deeply virtual Compton scattering}

\begin{figure}[!t]
\centering
\resizebox{0.65\columnwidth}{!}{\includegraphics{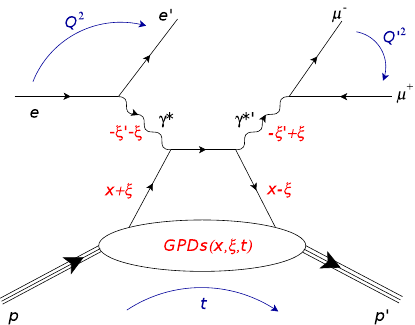}}
\caption{Schematic of the direct term of the DDVCS amplitude with a di-muon final state. The full amplitude contains also the crossed term where the final time-like photon is emitted from the initial quark. $Q^2$=-$q^2$ is the virtuality of the space-like initial photon, and $Q'^2$=$q'^2$ is the virtuality of the final time-like photon.}
\label{fig:ddvcs}
\end{figure}

There are essentially three experimental golden channels for direct measurements of the GPDs: the electroproduction of photons $eN \to eN \gamma$ which is sensitive to the DVCS amplitude, the photoproduction of a lepton pair $\gamma N \to Nl\bar{l}$ which is sensitive to the TCS (Timelike Compton Scattering)~\cite{Berger:2001xd,Boer:2015fwa} amplitude, and the electroproduction of a lepton pair $eN\rightarrow eNl\bar{l}$ which is sensitive to the DDVCS amplitude (Fig.~\ref{fig:ddvcs}). Only the latter provides the framework necessary for an uncorrelated measurement of the GPDs as a function of both scaling variables $x$ and $\xi$~\cite{PhysRevD.68.116005}. 

At leading twist and leading $\alpha_s$-order, the DDVCS process can be seen as the absorption of a space-like photon by a parton of the nucleon, followed by the quasi-instantaneous emission of a time-like photon by the same parton, which finally decays into a $l \bar{l}$-pair (Fig.~\ref{fig:ddvcs}). The scaling variables attached to this process are defined as 
\begin{eqnarray}
\xi' & = & \frac{Q^2-Q'^2+t/2}{2Q^2/x_\text{B}-Q^2-Q'^2+t} \label{xip_sca} \\
\xi  & = & \frac{Q^2+Q'^2}{2Q^2/x_\text{B}-Q^2-Q'^2+t} \label{xi_sca}
\end{eqnarray}
representing the Bjorken generalized variable ($\xi'$) and the skewness ($\xi$). If $Q'^2$=0, the final photon becomes real, corresponding to the DVCS process and leading to the restriction $\xi'$=$\xi$ in the Bjorken limit. If $Q^2$=0, the initial photon is real, referring to the TCS process and leading to the restriction $\xi'$=$-\xi$ in the Bjorken limit. In these respects the DDVCS process is a generalization of the DVCS and TCS processes. 

The DDVCS reaction amplitude is proportional to a combination of the CFFs ${\mathcal F}$ (with ${\mathcal F} \equiv \{ {\mathcal H}, {\mathcal E}, \widetilde{\mathcal H}, \widetilde{\mathcal E} \}$) defined from the GPDs $F$ (with $F \equiv \{ H,E,\widetilde{H},\widetilde{E} \}$) as
\begin{eqnarray}
\mathcal{F}(\xi',\xi,t) & = & \mathcal{P} \int_{-1}^1  F_+(x,\xi,t)\bigg[\frac{1}{x-\xi'} \pm \frac{1}{x+\xi'}\bigg] dx \nonumber \\
 & - & i \pi F_+(\xi',\xi,t) \label{eq1}
\end{eqnarray}
where ${\mathcal P}$ denotes the Cauchy's principal value integral, and  
\begin{equation}
F_+(x,\xi,t) = \sum_{q} \left(\frac{e_q}{e}\right)^2 {\left[ F^q(x,\xi,t) \mp F^q(-x,\xi,t) \right] }
\end{equation}
is the singlet GPD combination for the quark flavor $q$, where the upper sign holds for vector GPDs $(H^q,E^q)$ and the lower sign for axial vector GPDs $(\widetilde{H}^q,\widetilde{E}^q)$. In comparison to DVCS and TCS, the imaginary part of the DDVCS CFFs access the GPDs at $x$=$\pm \xi' \neq \xi$ instead of $\xi'$=$\pm \xi$, and the real part of the DDVCS CFFs involves a convolution with different parton propagators. Varying the virtuality of both incoming and outgoing photons changes the scaling variables $\xi'$ and $\xi$ and maps out the GPDs as function of its three arguments independently. From Eq.~\ref{xip_sca}-\ref{xi_sca}, one obtains 
\begin{equation}
\xi' = \xi  \, \frac{Q^2-Q'^2+t/2}{Q^2+Q'^2}
\end{equation} 
indicating that $\xi'$, and thus the imaginary parts of the CFFs $\{ \mathcal{H}, \mathcal{E} \}$, changes sign around $Q^2$=$Q'^2$. This represents a strong testing ground of the universality of the GPD formalism~\cite{Anikin:2017fwu}. 

\begin{figure}[!t]
\centering
\resizebox{0.75\columnwidth}{!}{\includegraphics{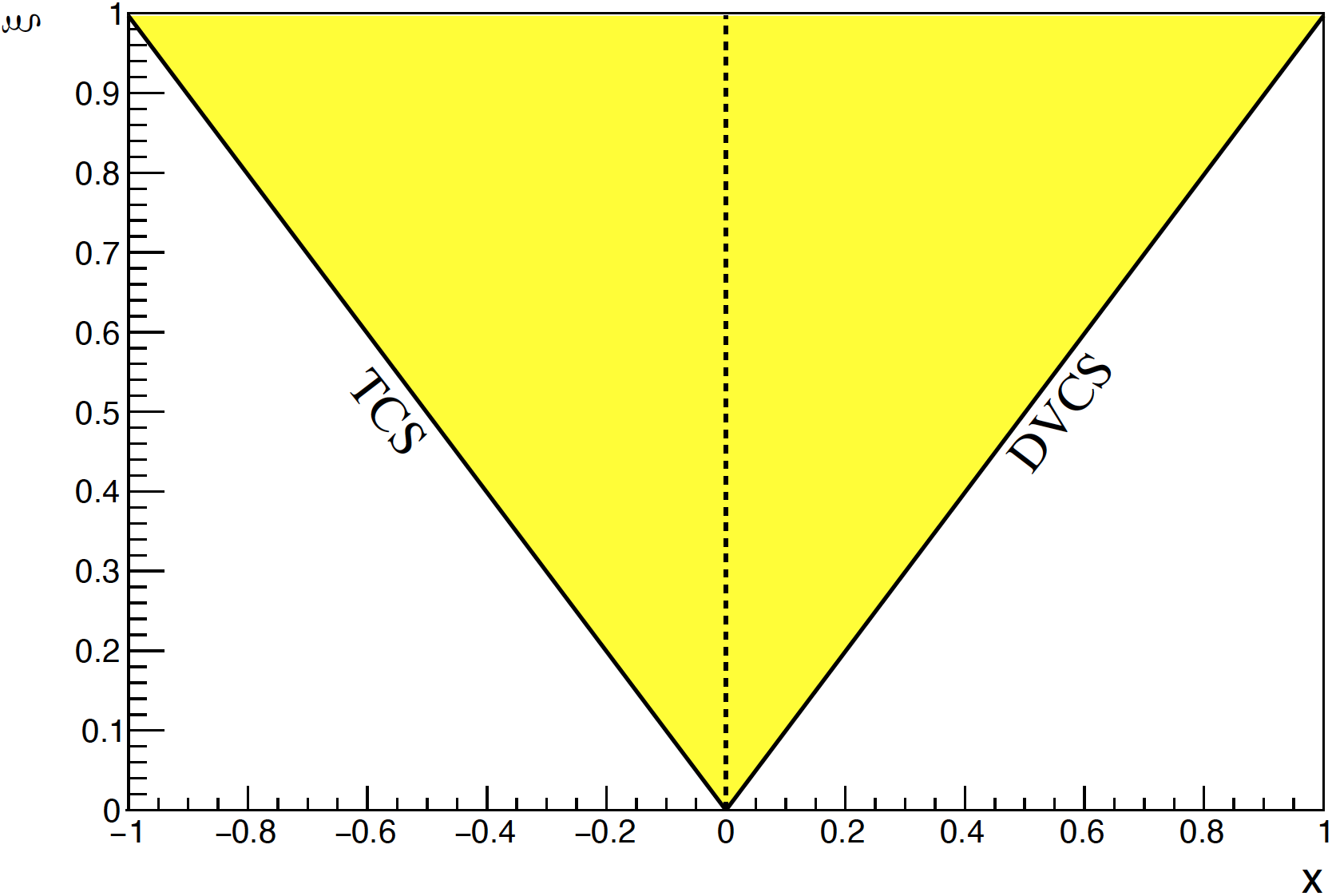}}
\caption{Singlet GPD $F_+(x,\xi,0)$ coverage of the physics phase-space from the imaginary part of the CFFs: the yellow area represents the DDVCS reach bounded on the one side by the TCS, and on the other side by DVCS lines. The $x$-axis corresponds to  the PDFs (Parton Distribution Functions) domain measured in inclusive Deep Inelastic Scattering.}
\label{Phy_Spa}
\end{figure}

As a consequence of the time-like nature of the final photon, the DDVCS process is restricted to the $\vert \xi' \vert < \xi$ region, and therefore the GPDs can be accessed only in the $\vert x \vert < \xi$ region (Fig~\ref{Phy_Spa}). Although the whole region $\vert x \vert > \xi$ is not accessed, this is a tremendous gain of information since no deconvolution is involved. Exploring the remaining part of the phase space, for instance to construct sum rules would need two space-like virtual photons~\cite{PhysRevLett.90.012001,PhysRevD.68.116005}. 

%
%

\section{Physics observables}

\begin{figure}[!t]
\centering
\resizebox{0.735\columnwidth}{!}{\includegraphics{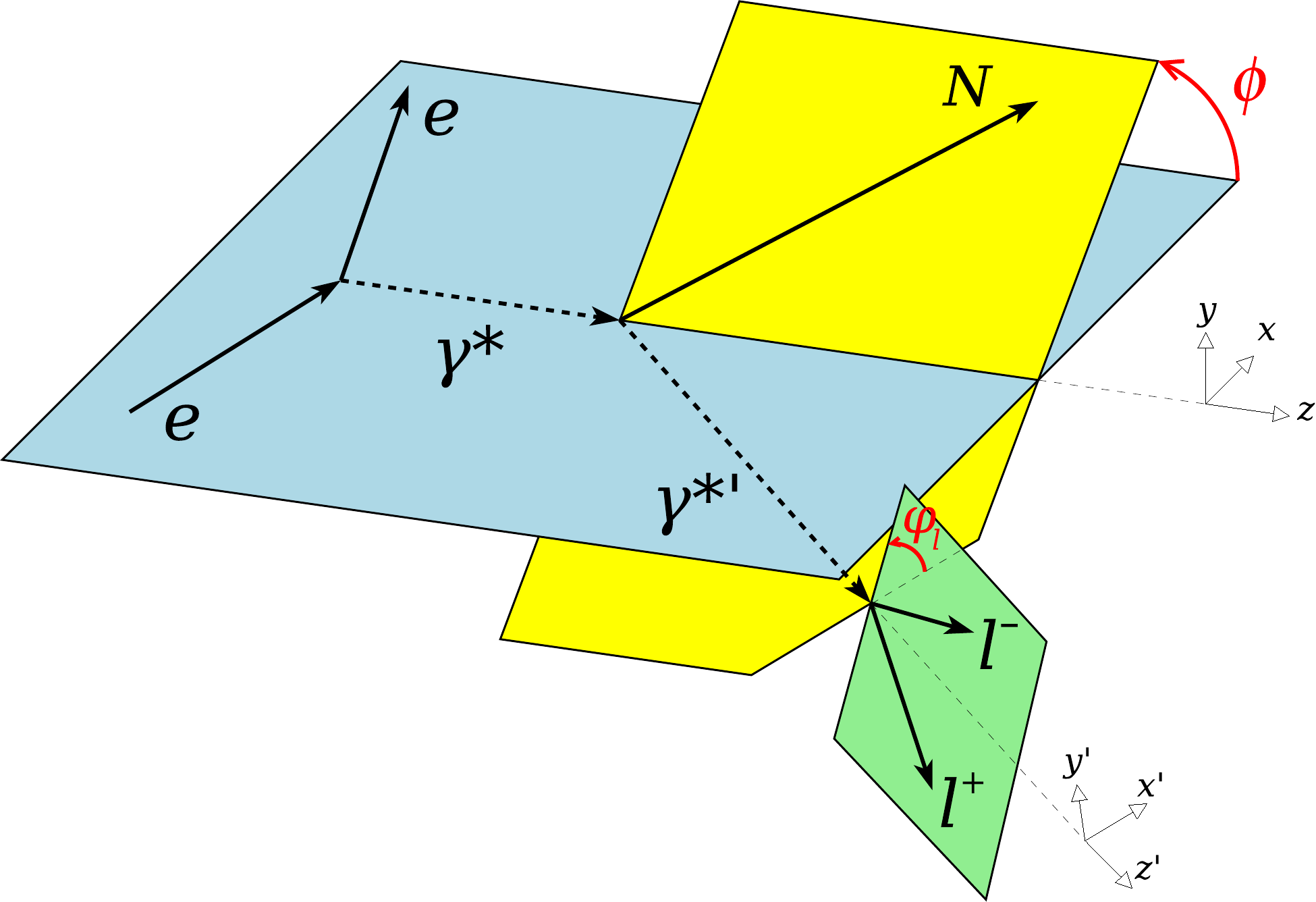}}
\caption{Reference frame of the $eN\rightarrow eNl\bar{l}$ reaction.}
\label{fig:frame}
\end{figure}

\begin{figure}[!t]
\centering
\resizebox{0.70\columnwidth}{!}{\includegraphics{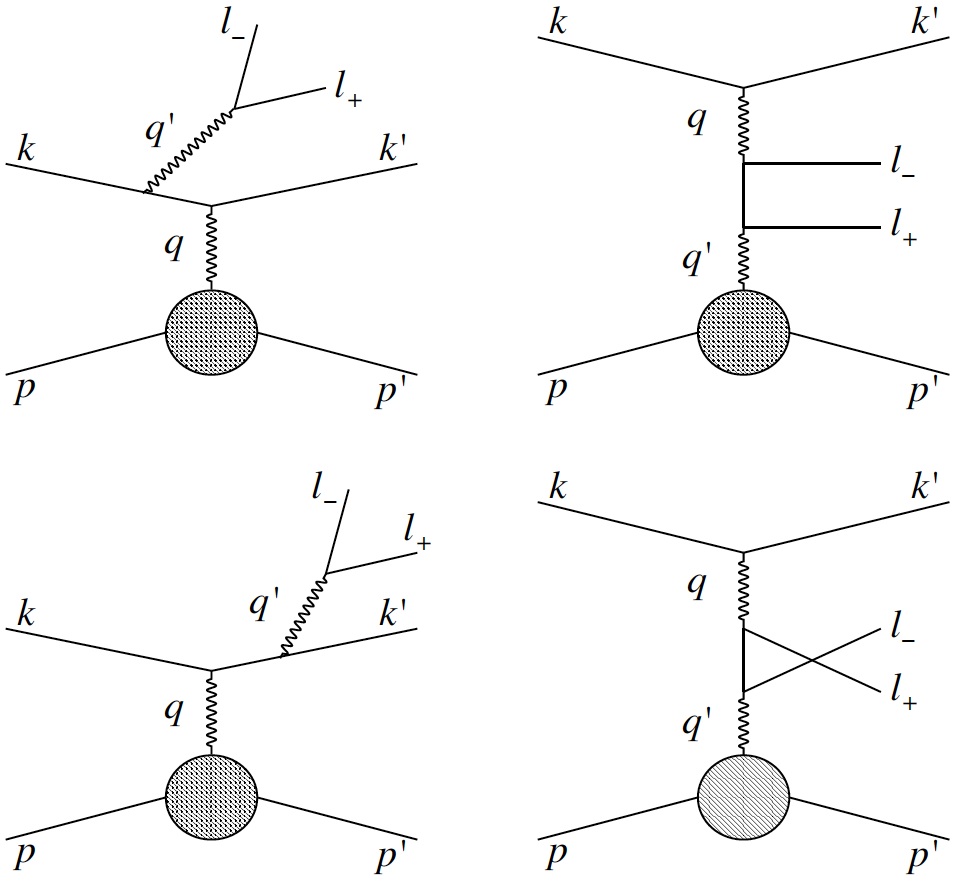}}
\caption{Different Bethe-Heitler processes contributing to the lepto-production of a di-muon pair besides DDVCS (Fig.~\ref{fig:ddvcs}), {\it i.e.} BH$_1$ (left) and BH$_2$ (right) as explained in the text.}
\label{fig:BH}
\end{figure}

The differential cross section of the lepto-production of a di-lepton pair has complicated kinematic dependences mixing the $(\phi,\varphi_l)$ out-of-plane angles in the angular distribution of di-leptons, as suggested by Fig.~\ref{fig:frame}. A further complexity in studying GPDs via DDVCS is the contribution of the Bethe-Heitler (BH) mechanisms to the same final state (Fig.~\ref{fig:BH}): the BH$_1$ process where the di-leptons are produced by time-like photons radiated by the incoming or outgoing electrons, and the BH$_2$ process where the di-leptons result from vacuum excitation within the nuclear field. The interference of these processes at the amplitude level result in complex kinematic dependences of the cross section, although the $BH$ amplitudes are precisely calculable at small momentum transfer $t$. Further difficulties may arise from antisymmetrization issues when the time-like photons decay into $e^+e^-$-pairs, but cancel out in the present study considering $\mu^+\mu^-$-pairs in the final state. The production of vector mesons subsequently decaying into a di-lepton pair remains a possible contamination source of the DDVCS  signal~\cite{PhysRevLett.90.012001}. Exploring the production of $\mu^+\mu^-$-pairs as function of the virtuality of the final photons gives a handle on the quantitative importance of this contamination.

\subsection{Cross section}

The lepton beam charge ($e$) and polarization ($\lambda$) dependence of the $eNl\bar{l}$ 7-fold differential cross section off unpolarized protons can be expressed as~\cite{PhysRevD.68.116005}
\begin{eqnarray}
d^7\sigma^{e}_{\lambda} & = & d^7\sigma_{BH_1} + d^7\sigma_{BH_2} + d^7\sigma_{DDVCS} + \lambda \, d^7\widetilde{\sigma}_{DDVCS}  \nonumber \\
 & - & e \, \left( d^7\sigma_{BH_{12}} + d^7\sigma_{INT_1} + \lambda \, d^7\widetilde{\sigma}_{INT_1} \right) \nonumber \\
 & + & d^7\sigma_{INT_2} + \lambda \, d^7\widetilde{\sigma}_{INT_2} 
\end{eqnarray}
with
\begin{equation}
d^7\sigma \equiv \frac{d^7\sigma} {dQ^2 \, dx_B \, dt \, dQ'^2 \, d\phi \, d\Omega_\mu} \, .
\end{equation}
In this expression, the index $BH_{12}$ denotes the interference amplitudes between the BH processes, and the index $INT_i$ denotes the $BH_i$-$DDVCS$ ones; $d^7\sigma_{A}$ $(A \equiv \{BH_1,BH_2,BH_{12},DDVCS,INT_1,INT_2\})$ represents the beam polarization independent contributions of the cross section, whereas $d^7\widetilde{\sigma}_{A}$ are the beam polarization dependent contributions. Integrating over the muon solid angle $d\Omega_{\mu}$ offers an understanding of the $eNl\bar{l}$ closer to the $eN\gamma$ reaction. In this integration, the interference contributions originating from the BH$_2$ mechanism vanish and the cross section becomes~\cite{Zhao:2020th} 
\begin{eqnarray}
d^5\sigma^{e}_{\lambda} & = & d^5\sigma_{BH_1} + d^5\sigma_{BH_2} + d^5\sigma_{DDVCS} + \lambda d^5\widetilde{\sigma}_{DDVCS} \nonumber \\
& - & e \, \left( d^5\sigma_{INT_1} + \lambda \, d^5\widetilde{\sigma}_{INT_1} \right) \, 
\label{eq_ch2_5fXsec_decomposition}
\end{eqnarray}
where the beam helicity-dependent DDVCS contribution arises at the twist-3 level. Considering a polarized electron beam, the unpolarized cross section can be measured as  
\begin{eqnarray}
\sigma^-_{0} & = & \frac{d^5\sigma^-_{+} + d^5\sigma^-_{-}}{2} \nonumber \\
 & = & d^5\sigma_{BH_1} + d^5\sigma_{BH_2} + d^5\sigma_{DDVCS} + d^5\sigma_{INT_1} \, ,
\end{eqnarray}
and the polarized cross section difference as
\begin{eqnarray}
\Delta \sigma^-_{\lambda} & = & \frac{d^5\sigma^-_{+} - d^5\sigma^-_{-}}{2} \nonumber \\
& = & \lambda \, \left[ d^5\widetilde{\sigma}_{DDVCS} +  d^5\widetilde{\sigma}_{INT_1} \right] \, , 
\end{eqnarray}
which involve a combination of the unknown $DDVCS$ and $INT_1$ amplitudes. Considering further a polarized positron beam, the comparison between electron and positron observables provides the charge sensitive parts of the cross section
\begin{eqnarray}
\Delta \sigma_{0}^C & = & \frac{\sigma^-_{0} - \sigma^+_{0}}{2} = d^5\sigma_{INT_1} \\
\Delta \sigma_{\lambda}^C & = & \frac{\Delta\sigma^-_{\lambda} - {\Delta\sigma^+_{\lambda}} }{2} = \lambda \, d^5\widetilde{\sigma}_{INT_1} 
\end{eqnarray} 
which isolate the interference amplitude, and the neutral observables
\begin{eqnarray}
\sigma_{0}^0 & = & \frac{\sigma^+_{0} + \sigma^-_{0}}{2} \nonumber \\
& = & d^5\sigma_{BH_1} + d^5\sigma_{BH_2} + d^5\sigma_{DDVCS} \\
\Delta \sigma_{\lambda}^0 & = & \frac{\Delta\sigma^+_{\lambda} + {\Delta\sigma^-_{\lambda}} }{2} = \lambda \, d^5\widetilde{\sigma}_{DDVCS} \label{eq:ddvcs0l}
\end{eqnarray} 
which select a $DDVCS$ signal. Similarly to DVCS~\cite{Voutier:2014kea}, combining observables obtained with polarized lepton beams of opposite charges allows the separation of the 4 unknown $INT$ and $DDVCS$ reaction amplitudes and permits an unambiguous access to combinations of the GPDs. In absence of positron beams, another possibility would be a Rosenbluth-like separation taking advantage of the different beam energy dependence of the $DDVCS$ and $INT_1$ amplitudes. However, the theoretical limitations reported for the DVCS channel~\cite{Defurne:2017paw} and the experimental complexity of the $eNl\bar{l}$ final state makes this method highly challenging for DDVCS. In practice, comparing electron and positron observables is the most reliable experimental technique for this separation. 

\subsection{Asymmetries}

Beam Spin Asymmetries (BSAs) offer an appealing access to the GPDs, experimentally easier to achieve than cross section measurements because of the cancellation of detector related effects, but at the expense of a more complex physics interpretation when the BH process does not dominate the unpolarized cross section. Considering polarized electron or positron beams, the BSA is expressed as 
\begin{eqnarray}
A_{LU}^{\pm}(\phi) & = & \frac{1}{\lambda^{\pm}} \, \frac{d^5\sigma_{+}^{\pm} - d^5\sigma_{-}^{\pm}}{d^5\sigma_{+}^{\pm} + d^5\sigma_{-}^{\pm}} \\ 
& = & \frac{d^5\widetilde{\sigma}_{DDVCS} \mp d^5\widetilde{\sigma}^\text{INT1}}{d^5\sigma_{BH_1} + d^5\sigma_{BH_2} + d^5\sigma_{DDVCS} \mp  d^5\sigma_{INT_1}} \nonumber
\end{eqnarray}
which shows a strict difference between electron and positron signals. Nevertheless, at twist-2 and in the $BH$-dominance hypothesis the electron and positron signals are just opposite in sign.

\begin{figure*}[t!]
\centering
\resizebox{0.497\textwidth}{!}{\includegraphics{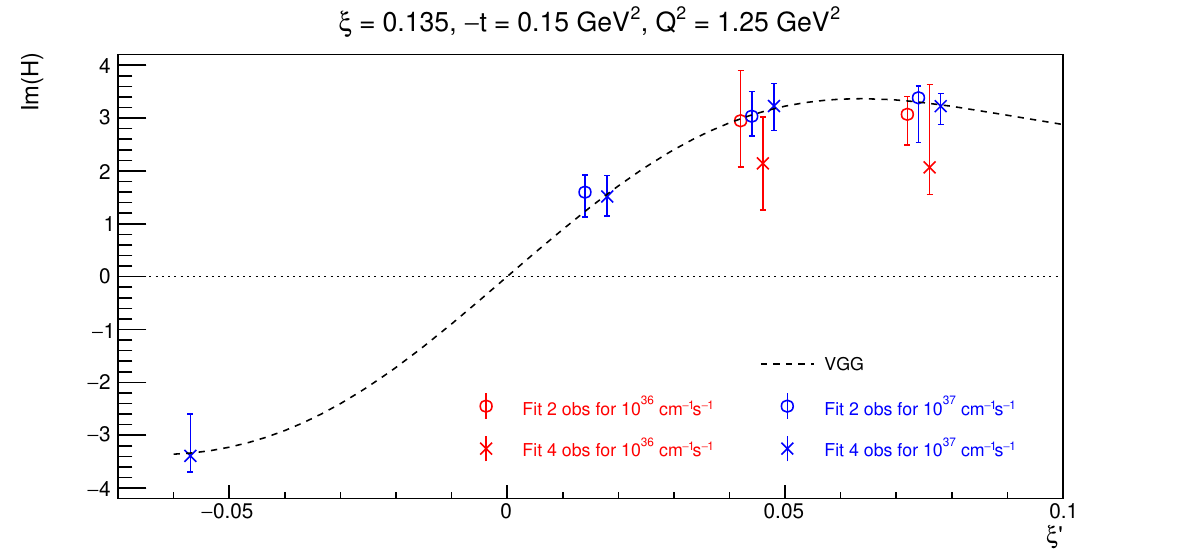}}
\resizebox{0.497\textwidth}{!}{\includegraphics{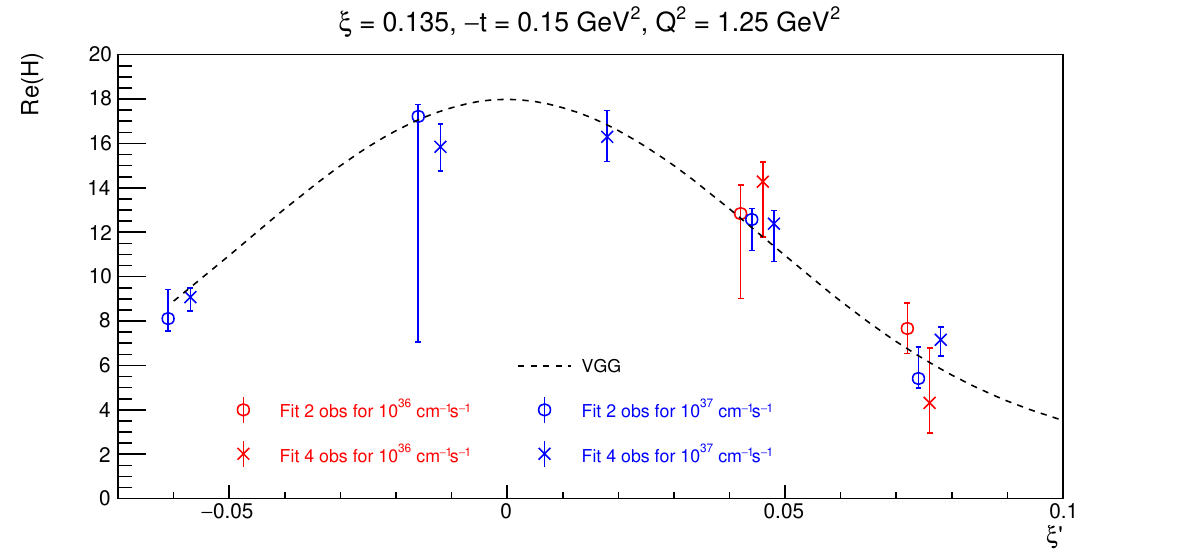}}
\caption{Imaginary (left) and real (right) parts of the CFF $\mathcal{H}$ extracted from the local fit of a set of 2 or 4 experimental observables, as explained in the text. The physics reach of the $10^{36}$~cm$^{-2}\cdot$s$^{-1}$ luminosity (red symbols) is compared to the $10^{37}$~cm$^{-2}\cdot$s$^{-1}$ one (blue points) at different $\xi'$-kinematics and a fixed $(\xi,t,Q^2)$ phase space point. The dashed lines represent the VGG model values used to generate pseudo-data. The data points defined in Tab.~\ref{TabKin}, are slightly offset in $\xi'$ for visual  clarity.}
\label{fig:CFF-H}
\end{figure*}

Experimentally, Beam Charge Asymmetries (BCAs) are more difficult observables to obtain than BSAs since some detector related effects (efficiency, solid angle) may still persist when comparing electron and positron data~\cite{Burkert:2020:pro}. The theoretical BCAs can be constructed as
\begin{eqnarray}
A_{UU}^{C}(\phi) & = & \frac{\left( d^5\sigma_{+}^+ + d^5\sigma_{-}^+ \right) - \left( d^5\sigma_{+}^- + d^5\sigma_{-}^- \right)}{d^5\sigma_{+}^+  + d^5\sigma_{-}^+ + d^5\sigma_{+}^- + d^5\sigma_{-}^- } \nonumber \\
& = & \frac{d^5\sigma_{INT_1}}{d^5\sigma_{BH_1} + d^5\sigma_{BH_2} + d^5\sigma_{DDVCS}}
\label{eq_ch2_obs_BCA} \\
A_{LU}^{C}(\phi) & = & \frac{ \left( d^5\sigma_{+}^- - d^5\sigma_{-}^- \right)/\lambda^- - \left( d^5\sigma_{+}^+ - d^5\sigma_{-}^+ \right)/\lambda^+}{d^5\sigma_{+}^+  + d^5\sigma_{-}^+ + d^5\sigma_{+}^- + d^5\sigma_{-}^- } \nonumber \\
& = & \frac{d^5\widetilde{\sigma}_{INT_1}}{d^5\sigma_{BH_1} + d^5\sigma_{BH_2} + d^5\sigma_{DDVCS} }
\label{eq_ch2_obs_BCAL}
\end{eqnarray}
which in the $BH$-dominance hypothesis constitute a pure interference signal. Additionally, the neutral BSA 
\begin{eqnarray}
A_{LU}^{0}(\phi) & = & \frac{\left( d^5\sigma_{+}^- - d^5\sigma_{-}^- \right)/\lambda^- + \left( d^5\sigma_{+}^+ - d^5\sigma_{-}^+ \right)/\lambda^+}{d^5\sigma_{+}^+  + d^5\sigma_{_-}^+ + d^5\sigma_{+}^- + d^5\sigma_{-}^- } \nonumber \\
& = & \frac{d^5\widetilde{\sigma}_{DDVCS}}{d^5\sigma_{BH_1} + d^5\sigma_{BH_2} + d^5\sigma_{DDVCS} } 
\label{eq_ch2_obs_BCA0}
\end{eqnarray}
quantifies the importance of higher twist effects.

\subsection{Compton Form Factors}

At leading twist and $\alpha_s$-order, the beam helicity-depen\-dent part of the $DDVCS$ amplitude vanishes and the $d^5 \sigma_{DDVCS}$ contribution to the cross section is proportional to the CFFs combination
\begin{eqnarray}
& {\mathcal F}_{DDVCS} = (1-\xi^2) {\left( {\mathcal H} {\mathcal H}^{\star} + \widetilde{\mathcal H} \widetilde{\mathcal H}^{\star} \right)} - \xi^2 \tau \widetilde{\mathcal E} \widetilde{\mathcal E}^{\star} \nonumber \\
& -\xi^2 {\left( {\mathcal H} {\mathcal E}^{\star} + {\mathcal E} {\mathcal H}^{\star} + \widetilde{\mathcal H} \widetilde{\mathcal E}^{\star} + \widetilde{\mathcal E} \widetilde{\mathcal H}^{\star}\right)} - {\left( \xi^2 + \tau \right)} {\mathcal E}{\mathcal E}^{\star} \, ,
\end{eqnarray}
while the beam helicity-independent part of the interference amplitude involves the real part of the combination
\begin{equation}
{\mathcal F}_{INT_1} = \frac{\xi'}{\xi} \left( F_1 {\mathcal H} - \tau F_2 {\mathcal E} \right) + \xi (F_1 + F_2) \widetilde{\mathcal H} \, , 
\end{equation}
and the beam helicity-dependent part involves the ima\-ginary part of the combination
\begin{equation}
{\mathcal F}'_{INT_1} = F_1 {\mathcal H} - \tau F_2 {\mathcal E} + \xi' (F_1 + F_2) \widetilde{\mathcal H} \, ,
\end{equation}
with $\tau$=$t/4M^2$~\cite{PhysRevD.68.116005}. The separation of the $INT_1$ amplitude allows to access a linear combination of CFFs, as compared to the bilinear combination of the $DDVCS$ amplitude. These combinations can be obtained unique\-ly as the $\phi$-moments of experimental observables. Note however that the CFFs content of $\phi$-moments depends on the approximations used to theoretically derive the cross section (leading twist, target mass corrections, higher twist, leading $\alpha_S$-order...)~\cite{Anikin:2017fwu}. 

The extraction of single CFFs from experimental data essentially falls into two groups: global~\cite{Kumericki:2009uq,Moutarde:2019tqa} and local~\cite{Guidal:2008ie,Kumericki:2011rz,Benali:2020vma} fits. On the one hand, the former involve the choice of a GPD parameterization and offer a coherent treatment of the measured and unexplored phase-spaces. However it leads to a model-dependent interpretation of experimental data which related systematic uncertainty is extremely difficult to evaluate~\cite{Moutarde:2019tqa}. On the other hand, the latter determine CFFs for each measured kinematic bin independently, are GPD model-independent, but suffer from the lack of uniqueness~\cite{Guidal:2008ie}. 

Following the fitter technique proposed to extract CFFs from DVCS observables~\cite{Guidal:2008ie}, 
a local fit method has been developed for  DDVCS~\cite{Zhao:2020th,Zhao:2020}. It consists in considering the eight CFFs (4 real parts and 4 imaginary parts) as free parameters of a simultaneous fit of the $\phi$-distributions of a set of experimental observables, using the VGG description of the $eN\mu^+\mu^-$ cross section. When the range of variation of the CFFs is limited, the dominant CFFs contributing to the fitted observables are obtained with reasonable error bars with respect to the fitting range.  Sub-dominant CFFs are generally not extracted by the fit because obtained with error bars corresponding to the limits of the variation domain, but they express their influence in the error of the dominant CFFs through correlations. These limits apart, this approach has been proved reliable and powerful in the DVCS case~\cite{Guidal:2008ie}. If only a restricted set of observables is available, limits and constraints can still be derived for specific  CFFs~\cite{Guidal:2009aa,Guidal:2010ig,Guidal:2010de}. Similar features have been observed for DDVCS~\cite{Zhao:2020th}. 

\begin{figure*}[t!]
\centering
\resizebox{0.70\textwidth}{!}{\includegraphics{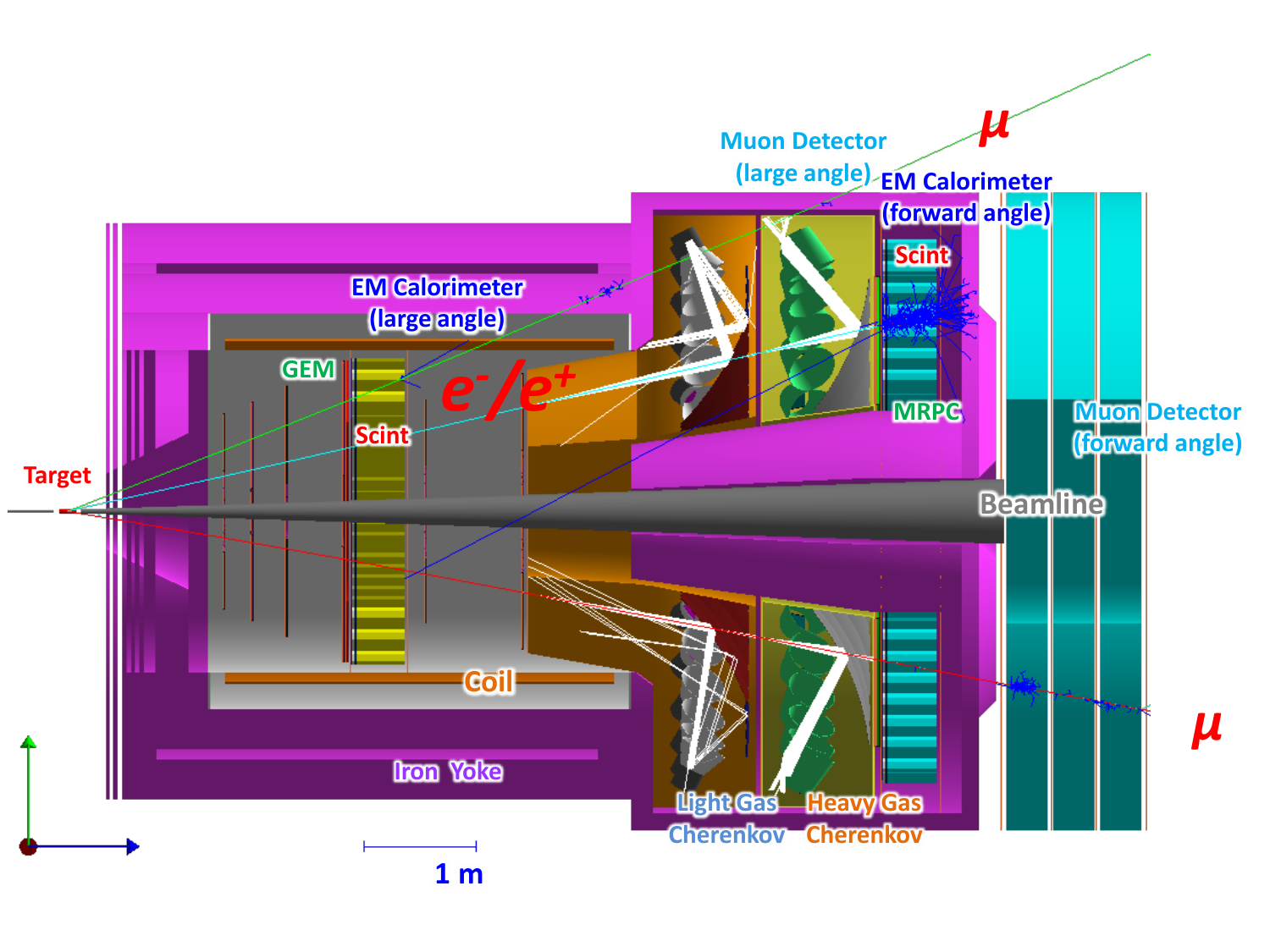}}
\caption{Schematic of the SoLID DDVCS setup in the Hall A of JLab supplemented with a muon detector at forward and large angles.}
\label{fig:solid}
\end{figure*}
Pseudo-data for the $(\sigma^\pm_0,\Delta\sigma^\pm_\lambda)$ observables are generated from VGG using proton GPDs compatible with current existing data~\cite{Dupre:2017hfs}, assuming an ideal detector and a beam energy of 11~GeV~\cite{Zhao:2019bzg}. The central values are smeared according to their statistical error bar determined for a data taking period of 50~days with each beam species 100\% polarized, and two luminosity scenarios $(10^{36},10^{37})$~cm$^{-2}\cdot$s$^{-1}$ for the kinematics and bin widths given in Tab.~\ref{TabKin}. The results of the fitting procedure for the CFF $\mathcal{H}$ are represented on Fig.~\ref{fig:CFF-H} for 5 selected $\xi'$-kinematics. Two experimental scenarios are considered: a 2-observables scenario $(\sigma^-_0,\Delta\sigma^-_\lambda)$ accessible with polarized electron beams, and a 4-observables scenario  $(\sigma^0_0,\Delta\sigma^C_0,\Delta\sigma^C_\lambda,\Delta\sigma^0_\lambda)$ requiring both polarized electron and positron beams. The real and imaginary parts of the CFF $\mathcal{H}$ are reported on Fig.~\ref{fig:CFF-H} whenever the fitting process delivers results with reasonable errors. They are further compared with the original CFF values used to generate pseudo-data (dashed line in Fig.~\ref{fig:CFF-H}). \newline
Independently of the scenario, the importance of high luminosity (blue versus red points) is striking, recovering all but one kinematics in the high luminosity case. The missing points in the vicinity of $\xi'$=$0$ feature small $\Delta\sigma^\pm_\lambda$ which make the fit very challenging. The 4-obser\-vables scenario tends to recover more kinematics than the 2-observables ones with a larger impact on the real part of $\mathcal{H}$ than on its imaginary part. This reflects the experimental access to a signal proportional to a bilinear or a linear CFFs combination. At leading twist, $\Delta \sigma^-_{\lambda}$ always accesses the imaginary part of a pure linear form (${\mathcal F}'_{INT_1}$) while it is only with the 4-observables scenario that $\Delta\sigma^C_0$ can access such a form (${\mathcal F}_{INT_1}$). As a consequence, the correlations between the fit extracted CFFs are weaker and error bars are reduced much beyond statistical expectations, a behaviour already observed in the DVCS channel~\cite{Burkert:2020:pro}. Note that while the qualitative features described above are somehow model-independent, their quantitative evaluation depends on the GPDs model used to generate pseudo-data. Nonetheless, the association of DDVCS detection capabilities and positron beams provides a unique mapping of the GPDs. 
\begin{table}[h!]
\centering
\resizebox{\columnwidth}{!}{\begin{tabular}{c|c|c|c|c|c|c|c|c}
\hline \hline
\multirow{2}{*}{$\xi'$} & \multirow{2}{*}{$\Delta\xi'$} & \multirow{2}{*}{$\xi$} & \multirow{2}{*}{$\Delta\xi$} & $Q^2$ & $\Delta Q^2$ & $-t$ & $\Delta t$ & $\Delta \phi $ \\
 & & & & \multicolumn{4}{c|}{(GeV$^2$)} & ($^\circ$) \\
\hline \hline
-0.060 & $\pm$0.030 & \multirow{5}{*}{0.135} & \multirow{5}{*}{$\pm$0.015} & \multirow{5}{*}{1.25} & \multirow{5}{*}{$\pm$0.25} & \multirow{5}{*}{-0.15} & \multirow{5}{*}{$\pm$0.05} & \multirow{5}{*}{$\pm$15} \\ \cline{1-2}
-0.015 & \multirow{4}{*}{$\pm$0.015}& & & & & & & \\ \cline{1-1}
 \phantom{-}0.015 & & & & & & & & \\ \cline{1-1}
 \phantom{-}0.045 & & & & & & & & \\ \cline{1-1}
 \phantom{-}0.075 & & & & & & & & \\ 
 \hline \hline
\end{tabular}}
\caption{Kinematics and bin widths considered for the generation of pseudo-data of the fitting study.}
\label{TabKin}
\end{table}

%
%

\section{Experimental configuration}

The Solenoidal Large Intensity Device (SoLID) is a brand new spectrometer device (Fig.~\ref{fig:solid}) to be installed in the Hall A of JLab to operate with initial electron beams up to 11~GeV~\cite{Chen:2014psa}. It is designed to use a solenoid field to sweep away low-energy background charged particles and allow operation at very high luminosities in an open geometry with full azimuthal coverage. Based on custom high rate and high radiation tolerant detectors, SoLID can carry out experiments using high intensity unpolarized or polarized electron beams and unpolarized or polarized targets. It consists of  two groups of sub-detectors: the Forward Angle Detector (FAD), and the Large Angle Detector (LAD). The FAD group covers the $8^\circ$-$16^\circ$ polar angle range and constitutes of several planes of Gas Electron Multipliers (GEM) for tracking, a light gas \v{C}erenkov (LGC) for $e/\pi$ separation, a heavy gas \v{C}erenkov (HGC) for $\pi/K$ separation, a Multi-gap Resistive Plate Chamber (MRPC) for time-of-flight measurement, a Scintillator Pad (SPD) for photon rejection and a Forward Angle Electromagnetic Calorimeter (FAEC). The LAD group covers the $16^\circ$-$28^\circ$ polar angle range and constitutes of  several planes of GEM for tracking, a SPD and a Large Angle Electromagnetic Calorimeter (LAEC). By reversing the polarity of the solenoid field, SoLID can operate similarly with initial positron beams and the field difference between different polarities is at 10$^{-5}$ level.

\begin{figure}[!t]
\centering
\resizebox{1.0\columnwidth}{!}{ \includegraphics{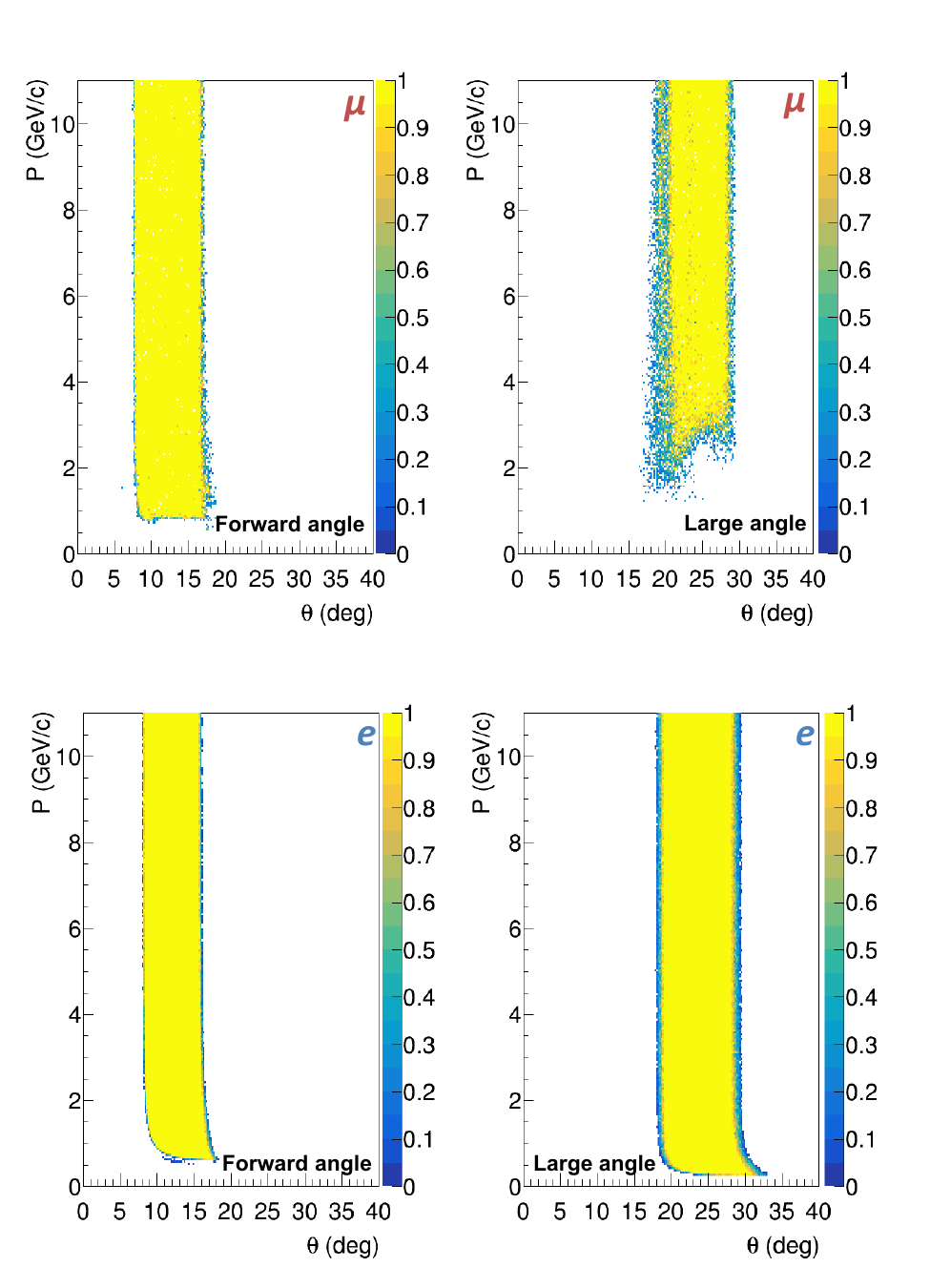}}
\caption{The acceptance for $\mu^-\mu^+$ (top) and $e^-/e^+$ (bottom) at the SoLID forward angle (left) and large angle (right) detectors.}
\label{fig:acc}
\end{figure}

Electrons and positrons are detected and identified by measuring their momenta, time-of-flight, produced photons in the threshold \v{C}erenkov detectors, and energy losses in the calorimeters. The SoLID spectrometer will be completed with a specific device dedicated for muon detection. The Large  Angle Muon Detector (LAMD) takes advantage of the material of the LAEC and the iron flux return to serve as shielding and a couple layers of GEMs at the outer radius of the downstream encap ensure the detection of particles. The Forward Angle Muon Detector (FAMD) placed after the downstream endcap consists of three layers of iron slabs instrumented with GEMs. The acceptance for muons and electrons/positrons according to the SoLID Geant4 detector simulation package is shown in Fig.~\ref{fig:acc}. 

The SoLID DDVCS experiment would ideally proceed by alternating electron and positron data taking periods for a total amount of 100~days equally shared between each beam species. The experiment would operate with a 15~cm long unpolarized liquid hydrogen  target at a luminosity of 1.2$\times$10$^{37}$~cm$^{-2}\cdot$s$^{-1}$, corresponding to a 3~$\mu$A beam intensity at 11~GeV. The electron  beam polarization is currently unlimited and considered 85\% in the present study, a routine value at JLab. Following the PEPPo technique~\cite{Abbott:2016hyf} for the production of polarized positrons, the positron beam polarization is instead correlated  with the beam intensity: the higher the intensity, the lower the polarization~\cite{Grames:2019:loi}. Considering the luminosity  requirement of the SoLID DDVCS experiment, present source simulations support a 30\% beam polarization~\cite{Car18}. An unprecedented amount of data would be collected and can be used for cross sections $(\sigma_0^{\pm},\Delta\sigma_\lambda^{\pm})$ and BSAs ($A_{LU}^{\pm}$) studies. Alternatively to cross section comparisons, BCA observables can also be constructed.

%
%

\begin{figure}[t!]
\centering
\vspace*{-6.5pt}
\resizebox{0.467\textwidth}{!}{\includegraphics{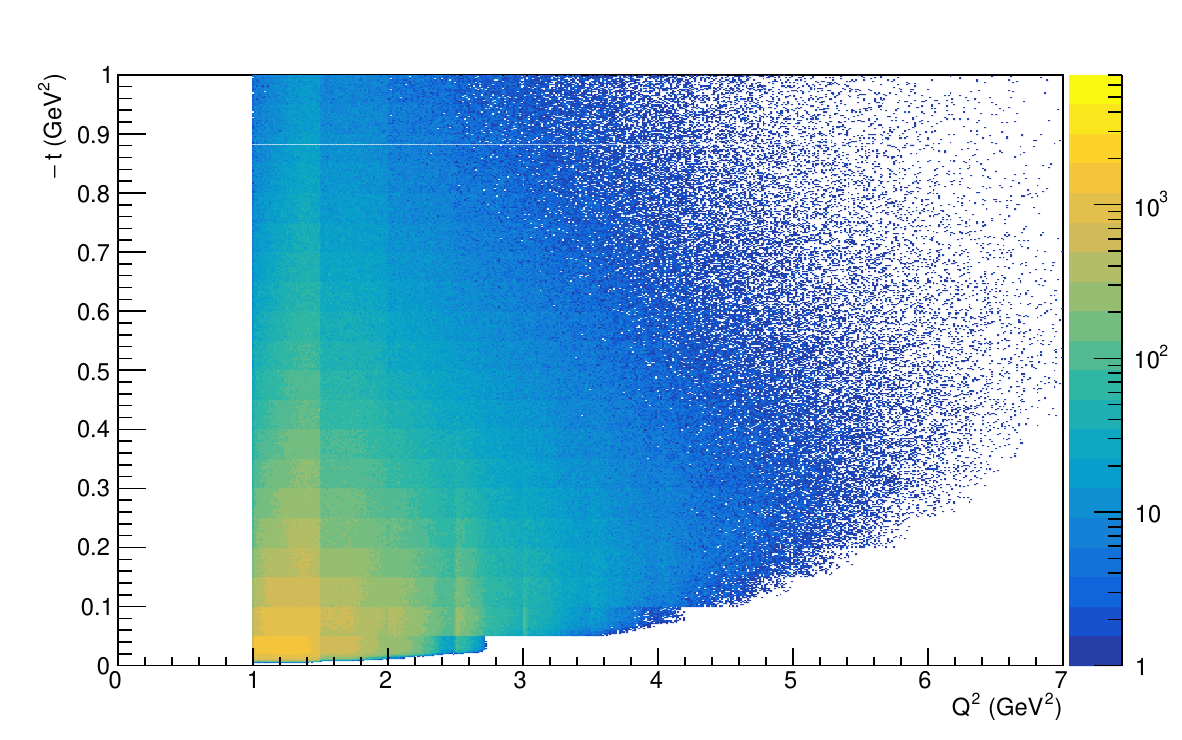}}
\resizebox{0.467\textwidth}{!}{\includegraphics{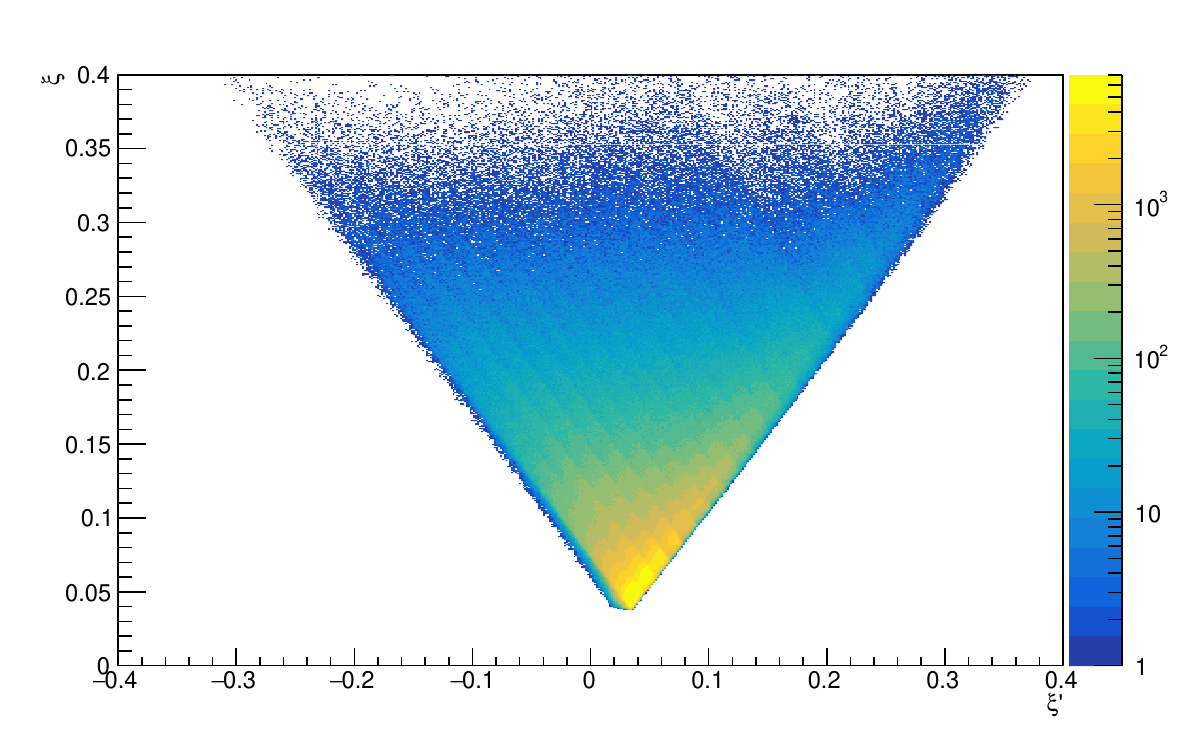}}
\caption{The distribution of the $eN\mu^+\mu^-$ event count number in the ($Q^2,-t$) plane (top) and the ($\xi',\xi$) plane (bottom). The kinematic coverage is limited by the factorization regime condition (-$t \ll Q^2$), and the DVCS ($\xi'$=$\xi$) and TCS ($\xi'$=-$\xi$) boundaries.}
\label{fig:kinematics}
\end{figure}

\section{Projected data}

A DDVCS event generator based on the $H$-only VGG  model~\cite{PhysRevLett.90.012001} at leading-twist, and extended to include the $\{E,\widetilde{H},\widetilde{E}\}$ GPDs of the nucleon~\cite{Zhao:2019bzg}, has been developed to evaluate projected data of the SoLID DDVCS experiment. The twist-3 $d^5\widetilde{\sigma}_{DDVCS}$ contribution to the $eN\mu^+\mu^-$ polarized cross section currently vanishes in this approach further detailed in Ref.~\cite{Zhao:2020th}. For the detection of the scattered electrons/positrons and the produced muon-pairs, the SoLID DDVCS acceptance is considered and an overall 50\% detector efficiency is taken into account to obtain the $eN\mu^+\mu^-$ event counts for the data taking scenario previously described.

\begin{figure}[t!]
\centering
\resizebox{1.0\columnwidth}{!}{\includegraphics{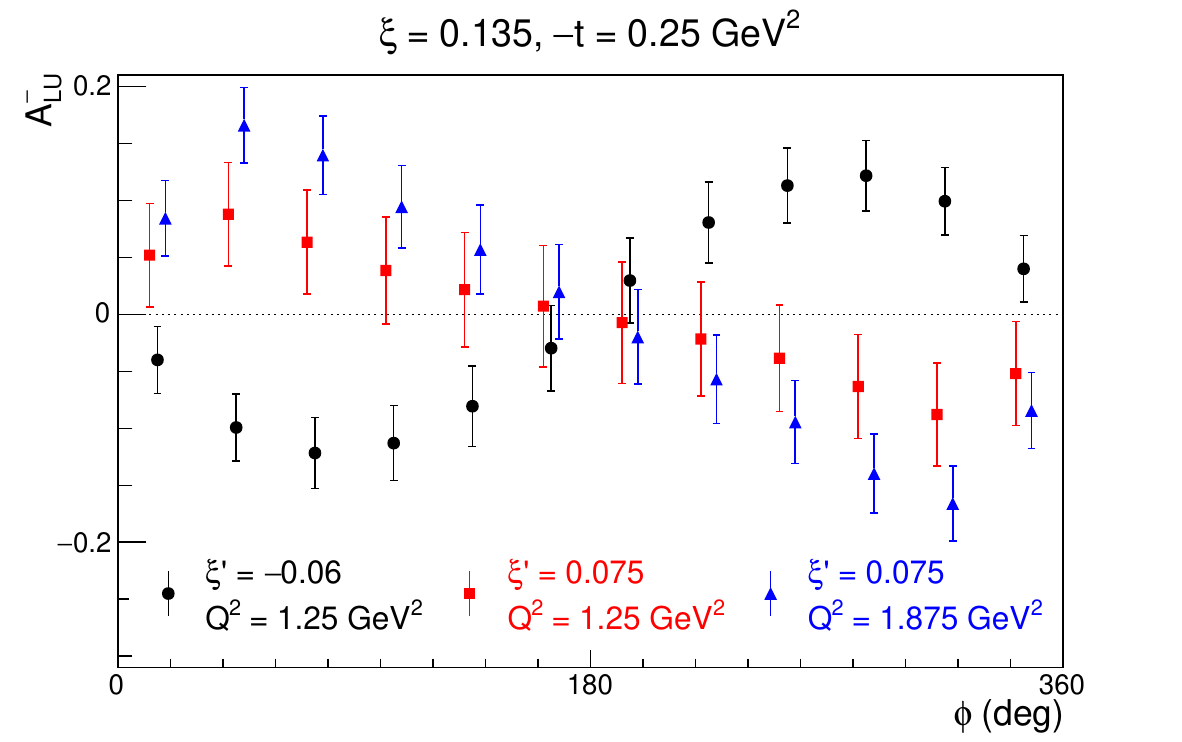}}
\resizebox{1.0\columnwidth}{!}{\includegraphics{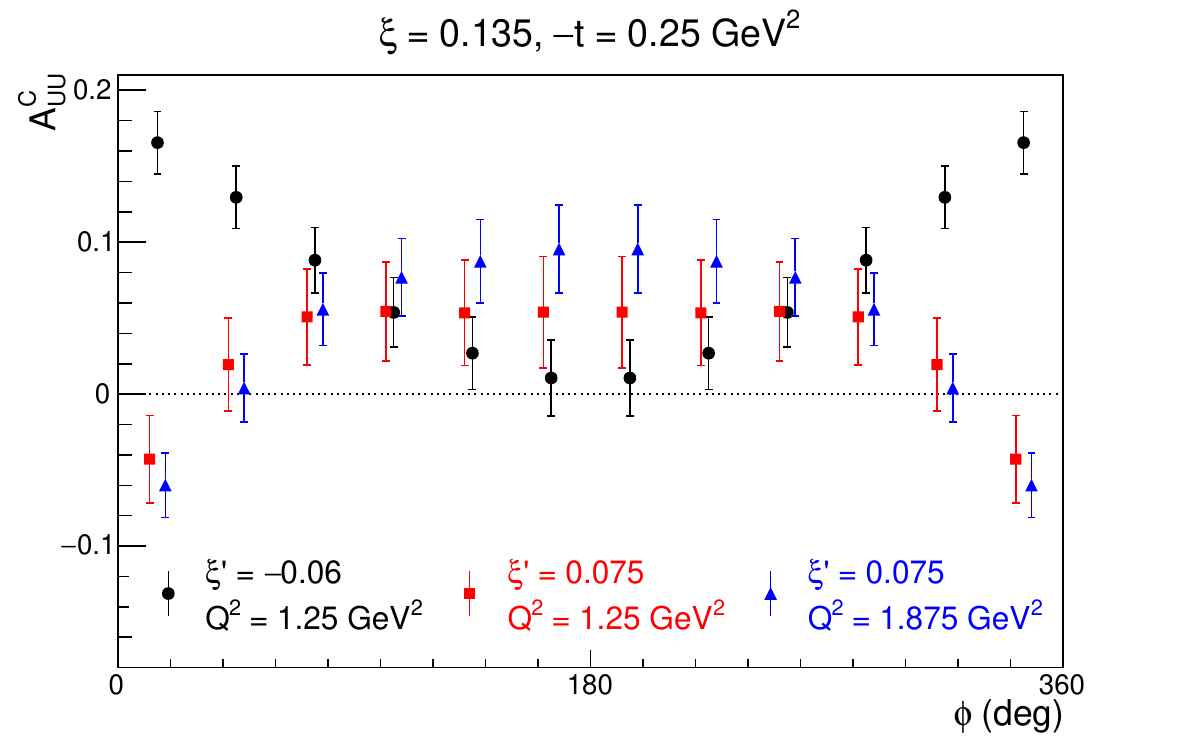}}
\caption{The electron BSA $A_\text{LU}^-$ (upper panel) and the unpolarized BCA ($A_{UU}^C$) (lower panel) for selected kinematics (Tab.~\ref{TabSta}) illustrating the $\xi'$ sign change.}
\label{fig:result}
\end{figure}

\begin{table}[b!]
\centering
\resizebox{\columnwidth}{!}{\begin{tabular}{c|c|c|c|c|c|c|c|c}
\hline \hline
\multirow{2}{*}{$\xi'$} & \multirow{2}{*}{$\Delta\xi'$} & \multirow{2}{*}{$\xi$} & \multirow{2}{*}{$\Delta\xi$} & $Q^2$ & $\Delta Q^2$ & $-t$ & $\Delta t$ & $\Delta \phi $ \\
 & & & & \multicolumn{4}{c|}{(GeV$^2$)} & ($^\circ$) \\
\hline \hline
-0.060 & $\pm$0.030 & \multirow{3}{*}{0.135} & \multirow{3}{*}{$\pm$0.015} & \multirow{2}{*}{1.250} & \multirow{2}{*}{$\pm$0.250} & \multirow{3}{*}{-0.25} & \multirow{3}{*}{$\pm$0.05} & \multirow{3}{*}{$\pm$15} \\ \cline{1-2}
\multirow{2}{*}{\phantom{-}0.075} & \multirow{2}{*}{$\pm$0.015}& & & & & & & \\ \cline{5-6}
 & & & & 1.875 & $\pm$0.375 & & & \\ 
 \hline \hline
\end{tabular}}
\caption{Kinematics and bin widths considered for SoLID projected data.}
\label{TabSta}
\end{table}
The distribution of the projected count number in the ($Q^2,-t$) and ($\xi',\xi$) physics phase-spaces are shown in Fig.~\ref{fig:kinematics}. The experiment covers a broad kinematic range limited on the one hand by the condition -$t \ll Q^2$ required for the factorization of hard and soft scale physics of the process, and on the other hand by the TCS and the DVCS correlation lines. With high statistics, the DDVCS reaction can be studied in all 5 kinematic variables  $(\xi',\xi,Q^2,t,\phi)$ independently. The full set of cross section and asymmetry observables can be measured over these  phase-spaces, allowing to constrain GPDs in uncharted territories. Examples of the $\phi$-distri\-butions of projected data are  shown in Fig.~\ref{fig:result} for the BSA measured with a polarized electron beam, and the unpolarized BCA measured with unpolarized electron and positron beams. Three different $(\xi',\xi,Q^2,t)$ kinematics are chosen to illustrate the experimental exploration of the sign change of the BSA as $\xi'$ cross over from negative to positive values. The shape of the unpolarized BCA is also changing as $\xi'$ evolves, reflecting the symmetry properties of GPDs. The expected experimental signals are of  significant amplitudes, enabling a meaningful extraction of the CFFs from the $\phi$-modulation of observables over the broad kinematic domain of the experiment. These data will provide invaluable constraints for a coherent determination of GPDs through global fit methods.

%
%

\section{Conclusions}

Using polarized electron and positron beams, the SoLID spectrometer supplemented with a muon detector allows us to investigate the partonic structure of the proton through the DDVCS process. This reaction is the only known process accessing the GPDs away of the diagonals $\xi'$=$\pm\xi$, {\it i.e.} providing proton structure information of prime importance for the nucleon tomography, the angular momentum sum rule, the distribution of nuclear forces, etc. Combining electron and positron observables, the SoLID DDVCS experiment isolates a pure $BH$-$DDVCS$ interference signal, and enables the extraction of the real and imaginary parts of the CFF $\mathcal{H}$ with extraordinary coverage and precision. It also makes possible to explore the importance of higher twist effects, and the validity of the BH-dominance hypothesis. Positron beams, both polarized and unpolarized, when combined with the power of SoLID's high luminosity and large acceptance capabilities, make the DDVCS reaction reaching its full potential for the study of the nucleon structure through GPDs.

%
%

\begin{acknowledgements}

This article is part of a project that has received funding from the European Union's Horizon 2020 research and innovation program under agreement STRONG - 2020 - No~824093. It is based upon work supported by the U.S. Department of Energy, Office of Science, Office of Nuclear Physics under contract DE-AC05-06OR23177 and DE-FG02-03ER41231, the China Scholarship Council, and the French Centre National de la Recherche Scientifique.

\end{acknowledgements}

%
%

\bibliographystyle{spphys}       
\bibliography{pDDVCS-BCA}

%
%

\end{document}